\documentclass[nato]{crckbked}

\usepackage{graphicx}
\usepackage{klucite}

\makeindex

\begin{document}
\newcommand{\beq}{\begin{equation}}
\newcommand{\eeq}{\end{equation}}
\newcommand{\beqn}{\begin{eqnarray}}
\newcommand{\eeqn}{\end{eqnarray}}
\newcommand{\bmath}{\begin{subequation}}
\newcommand{\emath}{\end{subequation}}
\numreferences

\begin{article}

\begin{opening}
\title{Quasiparticle undressing: a new route to collective effects in solids}

\author{J.E. \surname{Hirsch}\email{jhirsch@ucsd.edu}}

\institute{Department of Physics, University of California, San Diego\\
La Jolla, CA 92093-0319}

\runningauthor{Th. Author, A. Nother, Y. Etanotherone}

\begin{abstract}
The carriers of electric current in a metal are quasiparticles  dressed by electron-electron interactions, which
have a larger effective mass $m^*$ and
a smaller quasiparticle weight $z$ than non-interacting carriers.  If the momentum dependence of the self-energy can be neglected, the effective mass enhancement and quasiparticle weight of quasiparticles at the Fermi energy are 
simply related by $z=m/m^*$ ($m$=bare mass). We propose that both superconductivity and ferromagnetism in metals are driven by
quasiparticle 'undressing', i.e., that the correlations between quasiparticles that give rise to the collective state
are associated with an increase in $z$ and a corresponding decrease in $m^*$  of the carriers. Undressing gives rise to
lowering of kinetic energy, which provides the condensation energy for  the collective state.
In contrast, in conventional descriptions of superconductivity and ferromagnetism the transitions to these collective states result in $increase$ in kinetic energy of the carriers  and are driven by lowering of potential energy and exchange energy respectively.
\end{abstract}
\end{opening}

\section{Particles and quasiparticles}

Quasiparticles are 'dressed' bare particles, and they have a smaller quasiparticle weight and a larger
effective mass than bare particles. In several superconductors and ferromagnets of current interest
 there is experimental evidence hat quasiparticles  'undress', and resemble more
free particles, when correlations build up and the system orders. Associated with this, that the kinetic energy,
that is supposed to be optimal in the Fermi liquid normal state, decreases rather than increases in
the ordered state. This behavior is
counterintuitive, since in a normal Fermi liquid description it is expected that quasiparticles should become
further dressed and less like  free particles when they develop the correlations leading to the collective state, and that
they should $pay$, rather than $gain$, kinetic energy. Several 'unconventional' theories have been proposed to
explain these phenomena. Instead we propose here that  in fact quasiparticle
undressing is a unifying concept that can describe these collective effect in $both$ new and conventional
materials. The only difference is that only in the newer materials is the 
'undressing phenomenology' strong enough that it is easily seen in experiments.

 \section{Dressing and undressing}

 The single particle Green's function for an electron in an electronic energy
band is
\beq
G(k,\omega)=\frac{1}{\omega-\epsilon_k-\Sigma(k,\omega)}
\eeq
Expanding the self-energy $\Sigma(k,\omega)$ around $\omega=0$
\beq
\Sigma(k,\omega)=\Sigma(k,0)+\omega\frac{\partial\Sigma}{\partial \omega}
\eeq
the Green's function can be written as
\beq
G(k,\omega)=\frac{z_k}{\omega-\tilde{\epsilon}_k}+G'(k,\omega)
\eeq
where the first term is the quasiparticle part and the second term is the incoherent part, and the 'quasiparticle weight' $z_k$ is given by
\beq
z_k=[1-\partial \Sigma/\partial \omega]^{-1}
\eeq
If the momentum dependence of the self-energy can be
neglected,
\beq
\Sigma(k,\omega)\sim\Sigma(\omega)
\eeq
then we have simply  $z_k=z$ and $
\tilde{\epsilon}_k=z\epsilon_k $ 
which implies that the quasiparticle weight and effective mass $m^*$ are simply related by
\beq
\frac{m}{m^*}=z
\eeq
hence a highly dressed particle will have a small quasiparticle weight and a large effective mass. For the models of interest
in this paper the 'local approximation' Eq. (5) is reasonable.
Finally, the 'kinetic energy' of the system is defined by
\beq
E_{kin}=\sum_k n_k \epsilon_k
\eeq
where the occupation number $n_k$ is obtained from the single particle Green's function
\beq
n_k=\int_{-\infty}^\infty d\omega f(\omega)(-\frac{1}{\pi}ImG(k,\omega))
\eeq
with $f$ the Fermi function. The discontinuity of $n_k$ at the Fermi surface is  the quasiparticle weight $z_k$. If $z_k$ was to increase
for example as the temperature is lowered,   Eq. (7) predicts that the kinetic energy of the
system would decrease. The existence of this general relation between quasiparticle weight increase and kinetic energy lowering was
pointed out  by Norman et al\cite{norman}.

In the presence of interactions bare particles become dressed  with quasiparticle weight $z$ and
effective mass $m^*\sim1/z$. The residual weak interactions between these quasiparticles may cause the system
to undergo a transition to a collective state. In this paper we discuss a description of superconductivity and
ferromagnetism  that predicts that when the collective state develops quasiparticles undress, namely, that $z$ increases and $m^*$
 decreases. The resulting
kinetic energy lowering  provides the condensation energy stabilizing the ordered state.  

\section{Low energy effective Hamiltonians}

The low energy effective Hamiltonians in our theory can be derived from  the single band generalized Hubbard model
\beq
H=-\sum_{<ij>\sigma}t_{ij}(c_{i\sigma}^\dagger c_{j\sigma}+h.c.)+\sum_{ijkl\sigma\sigma'}
(ij|1/r|kl)c_{i\sigma}^\dagger c_{j\sigma'}^\dagger c_{l\sigma'} c_{k\sigma}
\eeq
where $(ij|1/r|kl)$ are matrix elements of the Coulomb interaction involving Wannier orbitals
at sites $i,j,k,l$. Keeping only two center integrals, we have shown that the off-diagonal interactions $(ii|1/r|ij)\equiv (\Delta t)_{ij}$ and 
$(ij|1/r|ji)=(ii|1/r|jj)\equiv J_{ij}$  lead to simple descriptions of superconductivity\cite{super} and ferromagnetism\cite{ferro} respectively, that have many features in common
with the phenomena seen in real materials. It should be pointed out however that these off-diagonal interactions are not simply calculable by computing
matrix elements of the Coulomb interaction between $fixed$ Wannier orbitals\cite{molecule}.

The Hamiltonian that gives rise to (hole) superconductivity is
\beq
H_{sc}=H_{ex Hub}+\sum_{<ij>\sigma}(\Delta t)_{ij}
(n_{i,-\sigma}+n_{j,-\sigma})[c_{i\sigma} ^\dagger c_{j\sigma}+h.c.]
\eeq
and the one giving rise to ferromagnetism is
\beq
H_{fm}=H_{ex Hub}+\frac{1}{2}\sum_{<ij>}J_{ij}[ \sum_\sigma( c_{i\sigma}^\dagger c_{j\sigma}+h.c.)]^2 
\eeq
The 'extended Hubbard' Hamiltonian $H_{ex Hub}$ includes the kinetic energy in Eq. (9) and the ordinary density-density Coulomb
interactions $U=(ii|1/r|ii)$, $V_{ij}=(ij|1/r|ij)$. The 'bond charge repulsion' term in Eq. (11) describes both exchange and pair hopping 
processes\cite{ferro}, arising from the matrix elements $(ij|1/r|ji)$ and $(ii|1/r|jj)$ respectively.

These 'off-diagonal' interactions lead to a decrease of the effective mass and
associated with it a decrease in the kinetic energy as the collective states develop. For ferromagnetism, the effective hopping
for a carrier of spin $\sigma$ is
\beq
t_{ij}^{eff}=t_{ij}-J_{ij}<c_{i,-\sigma}^\dagger c_{j,-\sigma}+h.c.>
\eeq
and it increases when spin polarization develops because the bond charge $<c_{i,-\sigma}^\dagger c_{j,-\sigma}>$
decreases. 
For superconductivity the effective hopping for an electron is
\beq
t_{ij}^{eff}=t_{ij}-n(\Delta t)_{ij}
\eeq
and it decreases monotonically as the number of electrons in the band increases. For holes instead the effective hopping amplitude is
\bmath
\beq
t_{ij}^{eff}=t^h_{ij}+n_h(\Delta t)_{ij}
\eeq
with
\beq
t^h_{ij}=t_{ij}-2(\Delta t)_{ij}
\eeq
\emath
and it increases as the hole concentration $n_h$ increases. Because the $local$ hole concentration around a given hole increases when holes
pair, the effective hopping Eq. (14a) will increase when pairing occurs\cite{london}.

These Hamiltonians describe changes in the quasiparticle effective mass when the system enters the collective state, and also as
function of doping in the normal state. However they do not properly describe the expected relation between effective mass and 
quasiparticle weight discussed in the previous section. This is most clearly seen for the effect of $\Delta t$ in the normal state.
According to the model Hamiltonian Eq. (10), the effective mass of a single hole in a filled band can be much larger than that of a single electron in an empty band, their ratio is
\beq
\frac{m^*_{hole}}{m^*_{electron}}=\frac{t}{t-2\Delta t}
\eeq
On the other hand, the quasiparticle weights for the single electron and the single hole in this single band model are simply
$z_{el}=z_{hole}=1$,
hence the expected relation between quasiparticle weight and effective mass Eq. (6) is strongly violated.

In addition, the optical conductivity sum rule is violated . The integral of the optical conductivity in a tight binding model is given by
\beq
\int_0^\infty d\omega \sigma_1(\omega) \sim <-T_{kin}>
\eeq
(proportionality factors are omitted).  For the Hamiltonian Eq. (10), 
the average kinetic energy for holes from the $ij$ bond is
\bmath
\beq
<T_{kin}>_{ij}=-(t_{ij}^h+n_h(\Delta t)_{ij})<c_{i\sigma}^\dagger c_{j\sigma}>+<T_{kin}^a>_{ij}
\eeq
\beq
<T_{kin}^a>_{ij}=-(\Delta t)_{ij}[<c_{i,-\sigma}^\dagger c_{i\sigma}^\dagger><c_{j\sigma} c_{i,-\sigma}>
+<c_{j,-\sigma}^\dagger c_{i\sigma}^\dagger><c_{j\sigma} c_{j,-\sigma}>]
\eeq
\emath
and it decreases below $T_c$ as the anomalous expectation values in Eq. 17(b) become nonzero. Hence the integrated
optical spectral weight (left side of Eq. (16) increases. A similar 
situation occurs for ferromagnetism.
However, in a real system the total optical spectral weight is conserved (optical sum rule), hence the optical sum rule is
'violated' if kinetic energy lowering occurs.  The resolution of both this violation and the unphysical relation between $m^*$ and $z$ arises from
consideration of other degrees of freedom not contained in the effective Hamiltonian Eq. (9).

\section{Relation between particle and quasiparticle operators}
The low energy effective Hamiltonian Eq. (9) should be understood as describing the dynamics of
quasiparticles rather than bare particles. For the description of superconductivity\cite{superundr} the quasiparticle operator
(which we denote by $\tilde{c}_{i\sigma}$) is related to the $coherent$ $part$ of the bare
particle operator $c_{i\sigma}$ by
\bmath
\beq
c_{i\sigma}=[1-(1-S)\tilde{n}_{i,-\sigma}]\tilde{c}_{i\sigma}
\eeq
with $0<S\leq 1$. Here, $c_{i\sigma}$ is an $electron$ operator. The corresponding relation for $hole$ operators is
\beq
c_{i\sigma}=S[1+\Upsilon \tilde{n}_{i,-\sigma}]\tilde{c}_{i\sigma}
\eeq
\emath
\beq
\Upsilon=\frac{1}{S}-1
\eeq
In the kinetic energy operator for bare electrons,
\beq
H_{kin}=-\sum_{<ij>\sigma}t_{ij}[c_{i\sigma}^\dagger c_{j\sigma}+h.c.]
\eeq
upon replacement of the bare electron operator in terms of the quasiparticle operators using Eq. (18), a 'correlated hopping' term of the form Eq. (10) results. Hence we can identify the correlated hopping amplitude  as
\beq
(\Delta t)_{ij}=t_{ij}S(1-S)=t_{ij}^h\Upsilon
\eeq
The quasiparticle weight for holes in a filled band is $z_h=S^2$. When $z_h$ is small, holes are heavily dressed in the normal state low hole concentration regime, and the 'undressing parameter' $\Upsilon$ is large. When the hole concentration $n_h$ increases,
the quasiparticle weight and the quasiparticle bandwidth increase according to the relations
\bmath
\beq  
 z=z_h(1+\Upsilon\frac{n_h}{2})^2
\eeq
\beq
D=D_h(1+\Upsilon\frac{n_h}{2})^2
\eeq
\emath
As a consequence, the system becomes increasingly more coherent as the hole concentration increases. 

The relation between bare particle and quasiparticle operator Eq. (18) gives rise to hole superconductivity when the parameter $\Upsilon$ is large, as well as 
to the 'undressing' phenomenology observed in experiments whereby quasiparticles undress both when the temperature decreases and the system becomes 
superconducting and when the hole concentration is increased in the normal state. This  behavior qualitatively shown in Figure 1 was predicted\cite{polaron}
\begin{figure}
\label{fig1}
\centerline{\includegraphics[width=7.5cm]{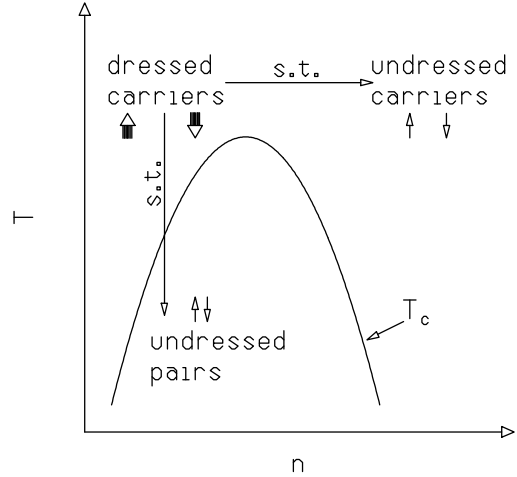}}

\caption{Undressing phenomenology in the cuprates. Both when hole carriers pair and when the hole concentration increases by doping,
there is an increase in the $local$ hole density around a given carrier that gives rise to undressing. This is accompanied by $s$pectral weight $t$ransfer 
from high to low frequencies.}
\end{figure}
based on microscopic considerations for the cuprates well before it was observed experimentally.

For ferromagnetism, a parallel analysis results when in the relation between bare and quasiparticle operators Eq. (18a) the site charge $n_{i\-\sigma}$
is replaced by the $bond$ charge. Replacement in the kinetic energy Eq. (20) gives rise to exchange and pair hopping terms that give rise to ferromagnetism 
and to undressing (increase in quasiparticle weight and decrease in quasiparticle mass) as the system develops spin polarization\cite{ferroundr}.

\section{Dynamic Hubbard models}
That the conventional single band Hubbard model is fundamentally flawed is seen as follows: in the Hubbard model,
destruction of an electron in a doubly occupied orbital yields the single electron state, ie
\beq
c_{i\sigma}|\uparrow \downarrow>=|-\sigma>
\eeq
This relation implies that the doubly occupied state $|\uparrow\downarrow>$ is a single Slater determinant. The very fact that electrons interact makes this an 
incorrect assumption. Hence the conventional Hubbard model fails to describe the most basic aspect of the electronic correlation problem it purports
to embody, namely correlation of electrons in the same Wannier orbital.

Recognition of the fact that the doubly occupied orbital $|\uparrow\downarrow>$ is a correlated state rather than a single Slater determinant
leads to dynamic Hubbard models\cite{dynhub}. The correct form of Eq. (23) is
\beq
c_{i\sigma}|\uparrow \downarrow>=|-\sigma>S+\sum_{n>0}|-\sigma>_nS_n
\eeq
where $|-\sigma>_n$ are excited state of the singly occupied orbital, and $|-\sigma>\equiv |-\sigma>_{n=0}$  the ground state.
Because electrons interact, $S$ in Eq. (24) is $never$ unity, and the second term on the right side of Eq. (24) is $never$ zero.
This leads to the relation between bare particle and quasiparticle
operators Eq. (18).

We have discussed various realizations of dynamic Hubbard models, involving either an auxiliary boson degree of freedom at each site
or more than one orbital per site\cite{dynhub1}. A new energy scale enters, given by the excitation energies of the states $|-\sigma>_n$. 
This is the energy range from which the high frequency spectral weight gets transfered from. In dynamic Hubbard 
models the Hubbard $U$ becomes a dynamical variable, which can take more than one value depending on the relative state of the
two electrons in the correlated state $|\uparrow \downarrow>$, and destruction of an electron in that correlated state $never$ yields the
singly occupied state $|\sigma>$ with its full amplitude. The study of dynamic Hubbard models is only in its beginning stages but it is
clear already that they exhibit very rich physics absent in the conventional Hubbard model.

\section{Conclusions and summary}

We are proposing that there is a single unifying concept behind the two most common collective effects in metals, superconductivity
and ferromagnetism: $quasiparticle$ $undressing$. Our proposal rests on four pillars, namely: (1) Theoretical consistency, (2) Experimental evidence,
(3) Microscopic justification, (4) Philosophical considerations. We summarize arguments in each category in the following.
\newline
\newline
{\bf (1) Theoretical consistency}

The theory is based on the relation beween bare particle and quasiparticle operators,
\beq
c_{i\sigma}^\dagger=[1-(1-S)\tilde{n}_{local}]\tilde{c}_{i\sigma}^\dagger + incoherent\ part
\eeq
with $S<1$. For the description of superconductivity and ferromagnetism $\tilde{n}_{local}$ is the site charge or the bond charge respectively (normalized
to unity). This relation
gives rise to low energy effective Hamiltonians with off-diagonal interaction terms $\Delta t$ and $J$ which drive hole superconductivity
and ferromagnetism respectively. Inclusion of both the local site and bond charge in Eq. (25) will yield a description of both
instabilities and their competition with the same Hamiltonian.

As the collective state develops, the term $\tilde{n}_{local}$ in Eq. (25) decreases: for ferromagnetism, spin polarization reduces the
bond charge density, and for superconductivity hole pairing reduces the electronic site charge density. As a consequence the
coefficient of $\tilde{c}_{i\sigma}^\dagger$ in the first term of Eq. (25) increases, and the incoherent part correspondingly decreases.
This leads to an increase in the quasiparticle weight and a decrease in the quasiparticle mass, as well as lowering of kinetic energy.
Spectral weight in the optical conductivity is transfered to low frequencies from the high frequency processes giving rise to
the incoherent terms in Eq. (25). This framework then maintains the simple relation between quasiparticle weight and quasiparticle
mass expected on general grounds, as well as explains the optical sum rule violation that occurs if only the low energy effective
Hamiltonians are considered. These low energy effective Hamiltonians are seen to give rise to ferromagnetism and superconductivity
both within mean field theory as well as in exact treatments.

For superconductivity, the theory predicts that it cannot occur when only electron carriers exist at the Fermi energy, because electron
carriers are already undressed and will not undress further by pairing. It can only occur when hole carriers exist at the Fermi energy
because holes are dressed. When the local hole concentration increases either by pairing or by hole doping in the
normal state, holes undress and turn into electrons. Pairing of holes is especially favored
when holes propagate in negatively charged structures, since in that case the dressing of the hole is highest and the undressing
associated with hole pairing is strongest\cite{molecule,dynhub}.
\newline
\newline
{\bf (2) Experimental evidence}

Certain ferromagnets exhibit clear evidence of undressing in optical spectroscopy: 
manganites\cite{okimoto}, $EuB_6$\cite{degiorgi}, $TlMn_2O_7$\cite{okamura}, and
some ferromagnetic semiconductors\cite{singley}. Furthermore the anomalous lowering of resistivity below $T_c$ and the negative
magnetoresistance observed in all ferromagnets may be interpreted as originating in lowering of effective mass upon spin
polarization.The undressing in optical spectroscopy may be too weak in certain metallic ferromagnets to be directly observed.
The undressing has not yet been seen in photoemission experiments in the manganites, but we expect that it will
be seen in the future. The reduction in the bond charge density associated with spin polarization that the
theory requires (see the previous subsection) manifests itself in the anomalous thermal expansion behavior of
ferromagnets below $T_c$\cite{janak}.

High $T_c$ cuprates exhibit abundant experimental evidence that quasiparticles undress when they pair, in photoemission\cite{ding}
and optical spectroscopy\cite{marel,santander}. Furthermore, optical\cite{uchida} and photoemission\cite{yusof} spectroscopy as well as transport\cite{ando} show that
holes undress in the normal state when the hole concentration increases by doping. The fact that by pairing dressed holes
turn into undressed electrons is seen directly in experiments in both conventional as well as high $T_c$ superconductors\cite{kik,hil,ver} .
The fact that superconductors are frequently  prone to lattice instabilities is naturally explained by the fact that many
$antibonding$ electronic states need to be occupied in order for the charge carriers to be hole-like. The observed isotope effects and
phonon signatures in tunneling are expected to result from the fact that ionic displacements  will affect the
magnitude of the interaction $\Delta t$ that drives superconductivity. The empirical observation  that superconductors have hole carriers
in the normal state was 
made by Chapnik\cite{chapnik}. 
\newline
\newline
{\bf (3) Microscopic justification}

Analysis of the physics of electrons in atomic orbitals shows that the basic relation Eq. (18) is
valid, with the parameter $S$ decreasing as the net ionic charge $Z$ decreases\cite{dynhub}. Also, first principles calculation of
hopping amplitudes in simple diatomic molecules show that indeed the hopping amplitude for holes is smaller than the
one for electrons in certain parameter ranges\cite{molecule}. The effect becomes large for small interatomic distance and
small $Z$, as expected. For ferromagnetism, first principles calculations of the relevant quantities have not yet been
reported.  
\newline
\newline
{\bf (4) Philosophical considerations}

According to the philosophical principle known as Occam's razor, 'pluralitas non est ponenda sine necessitas', 
plurality is not to be assumed without necessity.  If a single principle can
explain a variety of observations it should be preferred over multiple explanations. The principle of undressing provides a single
explanation for phenomena for which a very large number of different explanations have been proposed, hence it
should be preferred over the other explanations unless clearly proven wrong.

Another requirement on scientific theories is that they can be falsified by experiment. The present theory offers
plenty of opportunity for falsification. A few examples of possible observations that would  disprove the theory
are: finding of a single superconductor that does not have hole
carriers at the Fermi energy; observation of transfer of spectral weight in one or
two-particle spectral functions from $low$ to $high$ frequencies as the
collective state (superconductor or ferromagnet) develops; finding that in manganites the observed
effective mass reduction is $not$ accompanied by a substantially enhanced quasiparticle
peak in photoemission; finding that the gap slope in  superconductors does not have universal sign\cite{super};  etc.

\end{article}

\end{document}